\documentclass[%
 reprint,
bibnotes,
 amsmath,amssymb,
 aps,
prl,
]{revtex4-1}
\usepackage{caption}
\usepackage{subcaption}
\usepackage{graphicx}
\usepackage{dcolumn}
\usepackage{bm}

\begin{document}


\title{Interaction of hydrogen-edged boron nitride flakes with lithium: boron nitride as a protecting layer for a lithium-ion battery and a spin-dependent photon emission device}

\author{Narjes Kheirabadi $^1$, Azizollah Shafiekhani $^{1,2}$}
\affiliation{$^1$ Physics Department, Alzahra University, Vanak, Tehran 1993893973, Iran}%
\affiliation{$^2$ School of Physics, Institute for Research in Fundamental Sciences (IPM), P.O.Box:19395-5531, Tehran, Iran}%


\begin{abstract}

Abstract:
The current rechargeable battery technologies have a failure in their performance at high pressure and temperature. In this article, we have brought theoretical insights on using boron nitride flakes as a protecting layer for a lithium-ion battery device and extended its application for a spin-dependent photon emission device. Hence, the electronic properties of pristine and lithium-doped hydrogen-edged boron nitride flakes have been studied by the first principle density functional theory calculations. In this study, we have discussed the stability, adsorption energies, bond lengths, electronic gaps, frontier molecular orbitals, the density of states, charge distributions, and dipole moments of pristine and lithium hydrogen-edged doped boron nitride flakes. 
\end{abstract}

\maketitle

\section{Introduction}
After graphene discovery by Geim and Novoselove \cite{rise}, 2-dimensional (2D) materials have opened a new era for applications \cite{roadmap}. One of these 2D materials is boron nitride (BN). Experimentally, a sheet of heterogeneous boron nitride, a honeycomb structure analogous to graphene or white graphene, was successfully isolated \cite{14, 15, geim, hunting, lin2014fabrication}, and their Raman spectra and photoluminescence (PL) properties have been studied \cite{hunting, dress, wickramaratne2018monolayer, silly, lin2014fabrication}. In addition, thanks to liquid exfoliation, electron microscopy, Raman spectroscopy, and Fourier-transform infrared spectroscopy techniques, BN flakes are being produced and manipulate on a large scale \cite{han2008structure, zhi2009large, review, mendelson2019engineering, zhi2009large}.

BN nanomaterial has shown remarkable physical properties such as high thermal stability and conductivity (the second largest thermal conductivity per unit weight among all semiconductors and insulators), high resistance to oxidation, high electrical isolation, and outstanding chemical inertness \cite{cai2019high, o, 12, li2016atomically}. Because of excellent thermal and chemical stability, BN is a good candidate for high-temperature applications \cite{13}. Also, this material is a good candidate to make an insulator to encapsulate materials, especially other 2D materials, such as graphene and other nanomaterials with a large portion of atoms exposed on the surface \cite{review}. Moreover, it has significant potential to be used as a dielectric layer in high-performance 2D material based electronic and photonic devices \cite{sr}. Besides, its naturally passivated surface makes it possible to be integrated with photonic structures such as waveguides and cavities \cite{gap}. This wide gap material is also a good candidate to fabricate and control stable emitter with optically addressable spin states, including UV light emitters \cite{watanabe}. This key point is useful to make BN based spin sensing applications, quantum information technologies, \cite{spin} and bioimaging \cite{lei2015facile, xue2016hydrothermal, thangasamy2016supercritical} where controlling and manipulating light at the nanoscale is important \cite{kim2018photonic}.    

Advances made by this paper in the field of BN characteristics and applications include three main aspects. First, the interaction of BN flakes with lithium (Li) atom is still an open issue that we want to answer in this article and for the first time the adsorption of the Li atom on hydrogen-edged BN flakes is predicted. Indeed, the interaction of BN with alkali metals is a field of research \cite{potasium, addition, 16}. Interaction of infinite BN by Li atom has been studied \cite{16}. Based on this study, the calculated formation enthalpy of the considered forms of Li intercalated BN is found to be positive that eliminates BN interaction without externally supplied energy. In Ref. \cite{20}, Freeman and Larkindale have also discussed that BN powder reacts with sodium and potassium. Accordingly, the interaction of BN film with potassium is so that charge transfer between the potassium and BN bonds is very small. Whereas, no reaction was observed for the case of Li \cite{potasium}. Second, according to the next section, we have shown that the level of calculations is high enough to have a logical picture of interactions in the quantum level and those have also been endorsed by the vibrational frequencies calculations that make this article different from previous results. 
Third, in this article, we have suggested the application of BN flakes to improve the safety of Li-ion batteries and to use this material for photon emitters. The Li-ion battery is an attractive energy storage device exhibiting excellent efficiency of charge-discharge processes along with high energy density among the available other rechargeable batteries. Those are also the indispensable power source for a variety of portable electronic devices. Conventional rechargeable battery technologies that have an exceptional performance at ambient temperatures employ volatile electrolytes and soft separators, resulting in catastrophic failure under heat \cite{rodrigues2016hexagonal}. In combination with different conductive materials, h-BN exhibited improved electrochemical performance as the electrode, electrolyte, and separator in rechargeable batteries \cite{kumar2019review}. A stoichiometric mixture of hexagonal boron nitride, piperidinium-based ionic liquid, and a lithium salt electrolyte/separator extend the capability of Li-ion batteries to high temperatures \cite{rodrigues2016hexagonal}. Li-ion cells with h-BN incorporation also exhibit excellent performance and operational stability, especially at fast and ultra-fast charging rates \cite{ergen2020hexagonal}. Based on this study, we have suggested the BN based protracting layer for Li-ion batteries, which is a vacancy in the current applications of the BN in secondary batteries. Based on the results of this article, we suggest that the coating of BN on the current collectors of a Li-ion battery will increase the safety of a rechargeable battery. By comparison with recent experimental work that a current collector is made by sandwiching a polyimide embedded with triphenyl phosphate flame retardant between two super thin copper layers, we estimate that our suggested method is easier and more economical to be applied in large scales \cite{ye2020ultralight,kim2012synthesis}. 
We also discuss a detailed mechanism of Li defect in stable BN quantum dots, which is an open issue in the field of photon emitters, and we suggest BN based spin-dependent photon emission devices.     
\section{Structures and computational methods}
\begin{figure}[h]
   \centering   
   \includegraphics[scale=0.48]{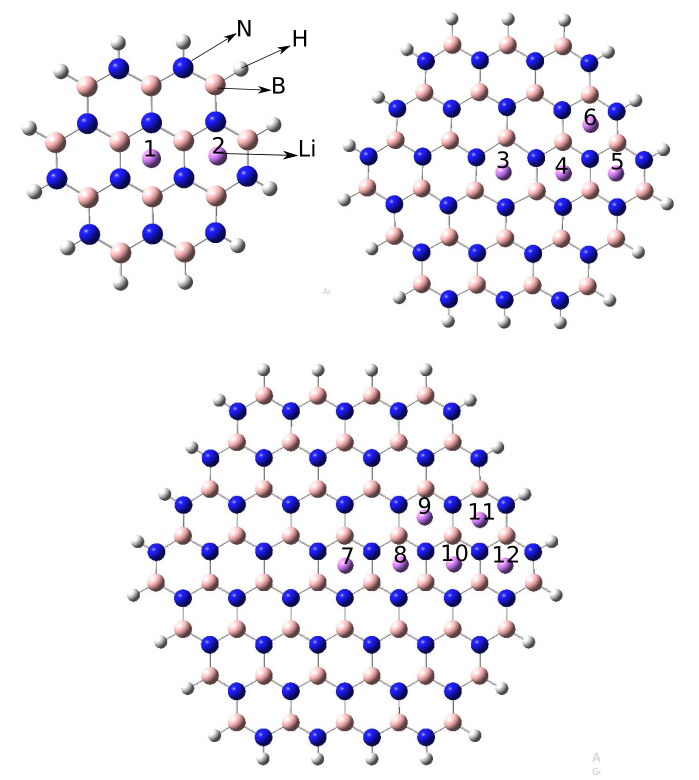}
   \caption [Models]{Li-doped flakes. Top left: two different Li (violet balls) doped $B_{12} H_{12} N_{12}$ systems. Top right: structures No. 3 to 6-2, $B_{27} H_{18} N_{27}$ related structures. Down part: 8 different doped $B_{36} H_{24} N_{36}$ flakes are depicted. The Li atom in each case has a different place, and the number of each structure is mentioned above the Li atom. Pink, blue and gray balls indicate the place of B, N, and H atoms, respectively.} 
    \label{5}
\end{figure}

The studied BN flakes are $B_{12} H_{12} N_{12}$, $B_{27} H_{18} N_{27}$ and $B_{36} H_{24} N_{36}$, and related Li-doped flakes. We have selected zigzag edge quasi-circular flakes because of experimental data that regularly observe hexagonal-shaped flakes with zigzag edges \cite{stehle2015synthesis, tay2014growth, dong2010boron}. The pristine and doped flakes are based on quasi-circular BN with hydrogen (H) atoms in the boundaries to saturate dangling bonds. Edge modification significantly enhances the stability of the flake. For example, in $Li B_{12} N_{12}$, the Li atom makes a covalent bond with an N atom in the edge. Hence, clean-edged BN flakes are not appropriate for rechargeable battery applications. Also, we predict that sulfurated and oxygenated flakes have more adsorption properties \cite{abdelsalam2020interaction}. Additionally, H atoms are unavoidable impurities that passivate the edges in the growth process of the BN flakes \cite{dong2010boron}. Each BN flake has been doped by one Li atom (Fig. \ref{5} shows these doped flakes, while the number of each structure is mentioned on the top of the Li atom in this figure). We have selected the Li atom because of its role in Li--ion battery applications. The Li atom is more electronegative compare to the other alkali metals; its electronegativity is equal to 1. Based on this electronegativity, we predict that Li-doped BN be more stable than other alkali metals doped BN. As an example, we consider flake No. 4 with one adsorbed sodium (Na) atom rather than one Li atom, the charge of adsorbed Na is -0.13 Mulliken, and its adsorption energy is -6.75 $KJ/mol$ (Na electronegativity is equal to 0.9). This amount is less than the adsorption energy of the flake No. 4 (-14.8 $KJ/mol$), Table. \ref{tab:table1}.

The method of the study is density functional
theory (DFT) method as implemented in Gaussian 03 \cite{g03}, which uses a Gaussian type of orbitals basis sets for electronic structure calculations. We have also used the Beke-Lee-Yang-Parr (B3LYP) exchange-correlation hybrid function \citep{b3lyp, lyp} and polarized 6-31g(d,p) basis set, in a situation where all symmetries of the initial structures are broken. Using this method, we expect to have a logical description of the electronic properties of doped BN flakes \citep{2dmaterials}.
  
To reach the optimized structure, first, by Gaussian 03, the BN flake has been geometrically optimized, and after that, a doping Li atom has been added to each flake. The initial state of the Li atom, as depicted in Fig. \ref{5}, is the center of a hexagon in the BN lattice. Hence, 15 different flakes are considered (Fig. \ref{5}). By this method, stability, the density of states (DOS), highest occupied molecular orbital-lowest unoccupied molecular orbital (HOMO-LUMO) gap, the distance of Li from the BN surface, frontier molecular orbitals (MOs) and charge distribution of each cluster have been studied. 

\section{Results}
\begin{figure}[h!]
   \centering   
   \includegraphics[scale=0.5]{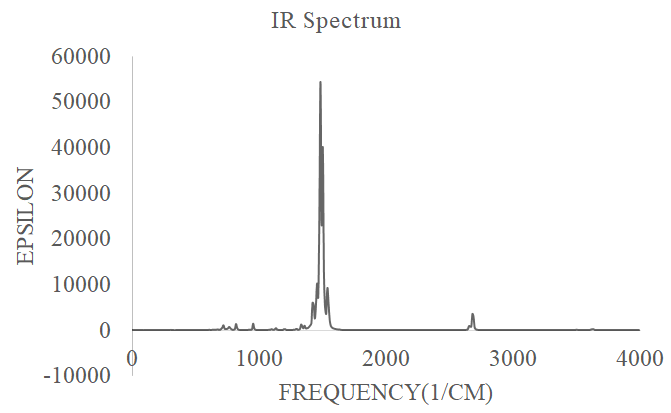}
   \caption {A typical IR spectrum of Li-doped flakes that is related to flakes No. 9-1 to 11-1.} 
    \label{IR}
\end{figure}

\begin{figure}[h]
   \centering   
   \includegraphics[scale=0.4]{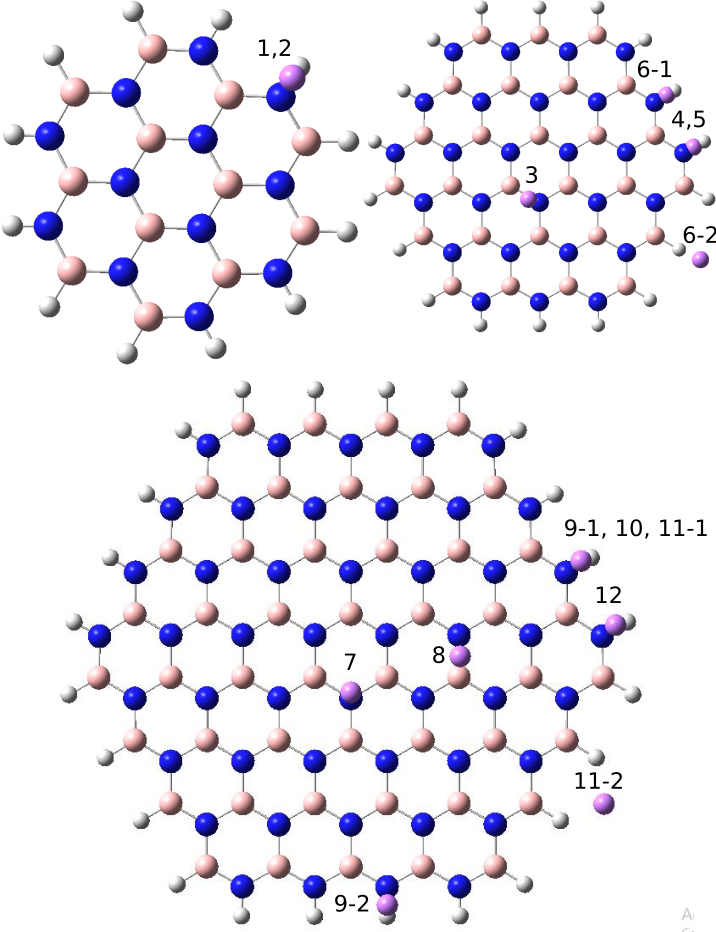}
   \caption [Models]{Results of optimized structures for flakes No. 1 to 12. This figure depicts the optimized Li atom position in each BN nano-flake. Pink, blue, gray, and violet balls indicate the place of B, N, H, and Li atoms, respectively.} 
    \label{results}
\end{figure}

The results of the calculations are tabulated in Table. \ref{tab:table1} and Fig. \ref{results}. Accordingly, there are ten possible locations as the result of the interaction of a Li atom with a BN flake (Fig. \ref{results}). However, these positions have different adsorption energies related to the different optimized positions of the Li atom. Note that $h$ in Table. \ref{tab:table1} is the distance of Li atom from a plane that passes through BN. Furthermore, all doped flakes have negative adsorption energy ($E_{ad}$); hence, in all cases, the adsorption has happened. Furthermore, the stability of all BN flakes, before and after adsorption, has been checked by vibrational calculations where all positive IR frequencies have revealed that BN flakes are stable for all vibrational modes (Fig. \ref{IR}). 

\widetext

\begin{table}[h!]
  \begin{center}
    \caption{Results of electronic properties of Li-doped BN flakes}
    \label{tab:table1}
    \begin{tabular}{c|c|c|c|c|c|c|c|c|c|c} 
      \textbf{No.} & \textbf{Chemical Name} & \textbf{$E_{ad}$} & \textbf{Li atom Charge} & \textbf{HOMO-LUMO Gap} & \textbf{h} &\textbf{Dipole Moment} & \textbf{Spin Polarization}\\
       &  & $(KJ/mol)$ & (Mulliken) & ($\alpha / \beta$) $(eV)$ & ($\AA$)& (Debye)& (Mulliken)\\
      \hline
      1 & $Li B_{12} H_{12} N_{12}$ & -13.4 & -0.22 & 2.0/6.4 & 2.1 & 5.3 & 0.02\\
      2 & $Li B_{12} H_{12} N_{12}$ & -13.4 & -0.22 & 2.0/6.4 & 2.1 & 5.3 & 0.02\\
      3 & $Li B_{27} H_{18} N_{27}$ & -1.5 & -0.04 & 2.9/5.3 & 4.0 & 0.8 & 0.01\\
      4 & $Li B_{27} H_{18} N_{27}$ & -14.8 & -0.23 & 2.0/6.0 & 2.1 & 5.7 &0.02\\
      5 & $Li B_{27} H_{18} N_{27}$ & -14.9 & -0.23 & 2.0/6.0 & 2.1 & 5.7 & 0.02 \\
      6-1 & $Li B_{27} H_{18} N_{27}$ & -19.1 & -0.24 & 2.2/6.0 & 2.2 & 5.8 & 0.02\\
      6-2 & $Li B_{27} H_{18} N_{27}$ & -3.7 & -0.11 & 2.3/5.9 & 1.8 & 3.8 & 0.01\\
      7 & $Li B_{36} H_{24} N_{36}$ &-1.4  & -0.04 & 2.9/5.2 & 4.0 & 0.8  & 0.01  \\
      8 & $Li B_{36} H_{24} N_{36}$ &-1.5 & -0.03 & 2.9/5.2  & 4.1 & 0.7 &0.01 \\
      9-1 & $Li B_{36} H_{24} N_{36}$ & -21.4 & -0.24 & 2.2/5.8 & 2.2 & 6.1 & 0.02 \\
      9-2 & $Li B_{36} H_{24} N_{36}$ &-21.2  &-0.24 & 2.2/5.8 &2.2 & 6.0 & 0.02 \\
      10& $Li B_{36} H_{24} N_{36}$ & -21.4 & -0.24 & 2.2/5.8 & 2.2 & 6.1 & 0.02\\
      11-1 & $Li B_{36} H_{24} N_{36}$ & -21.4 & -0.24 & 2.2/5.8 & 2.2 & 6.0 & 0.02 \\
      11-2 & $Li B_{36} H_{24} N_{36}$ & -3.7 & -0.14 & 2.1/6.0 & 1.6 & 4.7 & 0.04 \\
      12 & $Li B_{36} H_{24} N_{36}$ & -15.4 & -0.23 & 2.0/5.9 & 2.1 & 5.9 & 0.02\\
    \end{tabular}
  \end{center}
\end{table}

\endwidetext

\subsection{Stability, the Li distance from a BN surface and the length of bonds}
In the present work, we have attempted to locate the stable structures of pristine and Li-doped BN flakes. To do so, we have calculated the optimization energy of $B_{12} H_{12} N_{12}$, $B_{27} H_{18} N_{27}$ and $B_{36} H_{24} N_{36}$. $E_{ad}$ of these flakes are -2530359.2, -5679594.1, and -10084879.6 $KJ/mol$, respectively. By dividing the optimization energy of each flake into the total number of atoms, including equal numbers of N and B atoms, one can show that larger flakes are more stable than smaller ones. Additionally, this point is provable by another approach. If we assume that the hardness, $\eta$, of each flake is equal to $(E_{LUMO}-E_{HOMO})/2$, hardness of $H_{12} N_{12}$, $B_{27} H_{18} N_{27}$ and $B_{36} H_{24} N_{36}$ are 3.46, 3.29 and 3.21 eV, respectively. Furthermore, the chemical potential, $\mu=-(E_{LUMO}+E_{HOMO})/2$, of these three flakes are 3.28,  3.25, and 3.24 eV, respectively. Hence, the chemical stability of these flakes that is equal to $\mu^2/(2 \eta)$ can be calculated \cite{ouarrad2019size}. Accordingly, for pristine flakes, chemical stabilities are 1.56, 1.60, and 1.63 eV, respectively. Consequently, stability has increased by an increase in size. In comparison with the graphene flakes that by the increase of the size, their stability decreases \cite{vacancykhodam}, and this point is a keynote for the BN based quantum dot design.    

To calculate the $E_{ad}$ of each structure, we have subtracted the sum of the optimization energy of a Li atom and the optimization energy of a BN flake from the optimization energy of a doped BN \citep{likhodam}. In this way, we have calculated the $E_{ad}$ of each flake in Table. \ref{tab:table1}; the optimization energy of a Li atom is also -19667.6 $KJ/mol$.

According to Table. \ref{tab:table1}, all of the Li-doped structures are stable, while no chemical bonds arise. Besides, the most stable clusters are flakes No. 9-1, 9-2, 10 and 11-1. These clusters are between those clusters that have the largest BN flakes. For these flakes, the place of the adsorbed Li atom is on the center of a $N-H$ bond (Fig. \ref{results}). Then, cluster No. 6-1 is more stable. In this cluster, however the size of the structure is smaller, the adsorption happens above a $N-H$ bond. Next, clusters No. 12, 4, 5, 1, and 2 have a more absolute value of $E_{ad}$, respectively. These clusters have adsorbed a Li atom above a $N-H$ bond, but this bond is nearer to a B decorated edge. Afterwards, clusters No. 6-2 and 11-2 that have a Li atom near to a B decorated edge are stable. At last, those flakes that have the dopant in the center, means clusters No. 7, 8, and 3 have the lowest amount of $E_{ad}$, respectively. 

Consequently, the adsorption energy is, first, dependent on the place of the adsorbed Li, then it is dependent on the size of a cluster. If the size of a cluster is larger, the adsorption energy is larger, as well. According to Table. \ref{tab:table1}, those edges decorated by N atoms are the most stable places to accept a dopant. On the other hand, adsorption places in the center and at an edge decorated by B atoms are less stable. These calculations are done in zero Kelvin temperature. Considering thermal energy at room temperature equal to 2.5 $KJ/mol$ means that the clusters No. 3, 7, and 8 are not stable at room temperature. We predict that by the increase of the temperature from 0 Kelvin to room temperature, the adsorbed Li atom will receive enough energy to go to the edge \cite{25likhodam}.   

The distance of a Li atom from the BN layer for clusters No. 7, 8, and 3 is around 4 $\AA$. However, for those clusters that have a Li atom above a $N-H$ bond, the distance of the Li atom from the layer has decreased, drastically. So that according to Table. \ref{tab:table1}, this amount is around 2 $\AA$. For flakes No. 6-2 and 11-2 that have adsorbed the Li atom near to a $H-B$ decorated bond, theses amounts are 1.8 and 1.6 $\AA$, respectively. Consequently, the distance of a Li atom from the BN layer is dependent on its adsorption place and by being farther from the center, it decreases. 

Within each layer, B and N atoms are bonded by strong covalent bonds (Fig. \ref{results}), and the flat surface of BN includes $sp^2$ bonding between N and B atoms. The average length of a central $N-B$ bond for $B_{12} H_{12} N_{12}$, $B_{27} H_{18} N_{27}$ and $B_{36} H_{24} N_{36}$ is 1.45 $\AA$ comparable with 1.44 $\AA$ Reported in Ref. \cite{li2011large} for an infinite sheet. For $B_{12} H_{12} N_{12}$, $H-B$ and $H-N$, bond lengths are 1.19 and 1.01 $\AA$, respectively. For $B_{27} H_{18} N_{27}$ and $B_{36} H_{24} N_{36}$, these amounts are similar to the previous structure. Consequently, the increase of the size does not change the bond lengths of pristine flakes. Additionally, for all of the doped BN flakes, the average length of a chemical bond is similar to a pristine flake in order of $10^{-2}$ $\AA$. However, for clusters No. 6-2 and 11-2, the length of a $H-B$ bond, that is near to the adsorbed Li atom, has increased $0.01$ $\AA$. For clusters No. 1, 2, 4, 5, 6-1, 9-1, 9-2, 10 and 11-1 the length of a $H-N$ bond near to a Li atom also increases $0.01$ $\AA$ (Fig. \ref{results}). For instance, in the case of flakes No. 1 and No. 2, where the Li atom locates in the center of the nearest $H-N$ bond, the length of this bond is 1.02 $\AA$; while, before the adsorption of the Li atom, it is 1.01 $\AA$ (Fig. \ref{results} and Table. \ref{tab:table1}).

\subsection{DOS, HOMO-LUMO gap and frontier molecular orbitals}
\begin{figure}[h!]
   \centering   
   \includegraphics[scale=0.65]{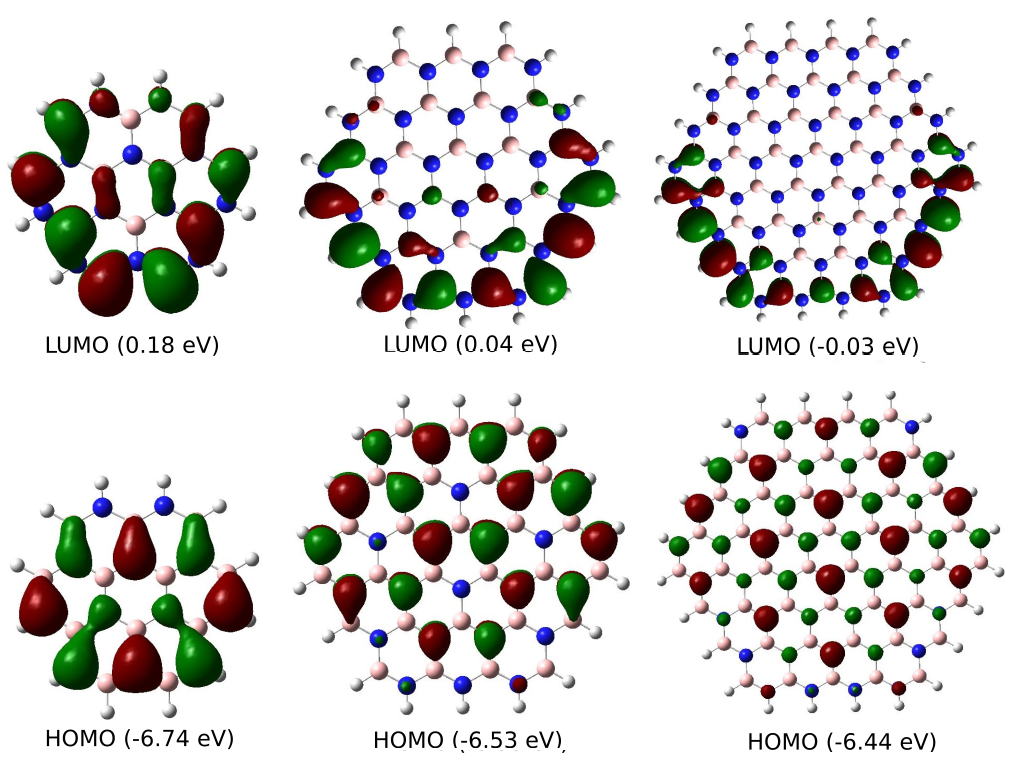}
   \caption [Models]{HOMO and LUMO orbitals for pristine BN flakes. The red and green colors stand for the positive and negative signs of the molecular orbital wave function, respectively. The isovalue surface is 0.02 $e/(Bohr)^3$.} 
    \label{MO1}
    \end{figure}

In this section, DOS and HOMO-LUMO gap are explored to study the reactivity of BN flakes; a larger HOMO-LUMO gap means larger excitation energy. Electronic gaps of $B_{12} H_{12} N_{12}$, $B_{27} H_{18} N_{27}$ and $B_{36} H_{24} N_{36}$, pristine BN flakes, are $6.9$, $6.5$ and $6.4$ eV, respectively. These values are comparable with the experimental evidence that seems to agree on a value near 6 eV for a BN sheet \cite{gap}. Considering a logarithmic behavior for the decrease of the gap by the size deduces that the limit of the experimental gap of a sheet (6 eV) happens for around 170 total number of N and B atoms. Consequently, pristine BN flakes are the wide gap insulators, and the gap increases as the size decreases because of the quantum confinement. Besides, by the increase of the size, the HOMO state has a redshift, and the LUMO state has a blueshift (Fig. \ref{MO1}) to decrease the gap. 

\widetext

\begin{figure}[h]
   \centering   
   \includegraphics[scale=0.6]{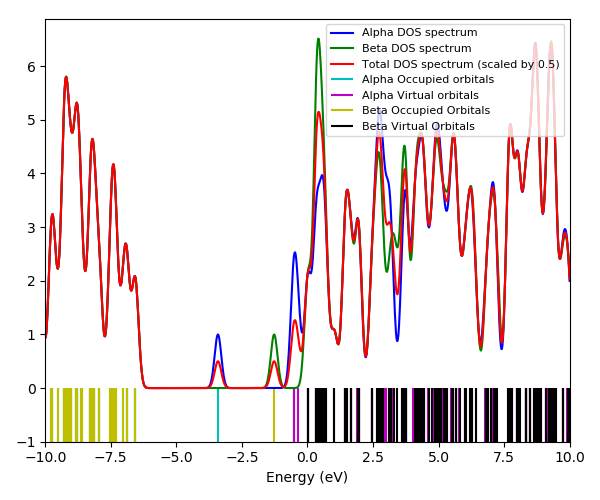}
   \caption [Models]{DOS spectrum of structure No. 3. The full width at half-maximum (FWHM) is 0.3 eV.} 
    \label{3}
\end{figure}

\endwidetext
Before, it has been shown that the absorption of a Li atom makes a graphene flake to be spin-polarized \cite{likhodam, spinkhodam, spinkhodam2}. In this work, we have also spin polarization of frontier MOs of Li-doped flakes. The only difference is that in the case of BN flakes, the last occupied states ($\alpha$ or $\beta$) and all virtual states are spin-polarized, and other occupied states are not spin-polarized. However, in the case of graphene, all of the occupied, and virtual states are spin-polarized; the DOS spectrum of the flake No. 3 is depicted in Fig. \ref{3}. For all of the understudy cases, the total Mulliken atomic spin densities are equal to 1. Hence, to calculate the spin polarization of the BN surface, we have calculated 1 - (spin density of Li atom). Based on this point, the Mulliken analysis shows that a little spin is transferred to the doped BN layer (Table. \ref{tab:table1}). 
About the gap of a Li-doped BN, first, the gap is dependent on the place of the Li atom related to the edge, Table. \ref{tab:table1}. Those flakes with a Li adsorbed on the edge, have a lower $\alpha$ gap, but a higher $\beta$ gap. And, if a cluster has a lower $\alpha$ gap, it will have a higher $\beta$ gap. These points are important for gap engineering. The highest $\alpha$ gap is 2.9 for clusters No. 3, 7, and 8; these BN flakes have a Li atom located in the center. The lowest $\alpha$ gaps belong to clusters No. 1, 2, 4, 5, and 12. These structures all have a Li atom on a $N-H$ bond near to a B decorated edge. For $\beta$ gap, the highest $\beta$ gaps belong to No. 1 and 2. The lowest gaps belong to clusters No. 7, 8, and then cluster No. 3; all of these clusters have a Li atom located in the center of the cluster, and the adsorbed Li is not near to the edge states. 

\begin{figure}[h]
   \centering   
   \includegraphics[scale=0.47]{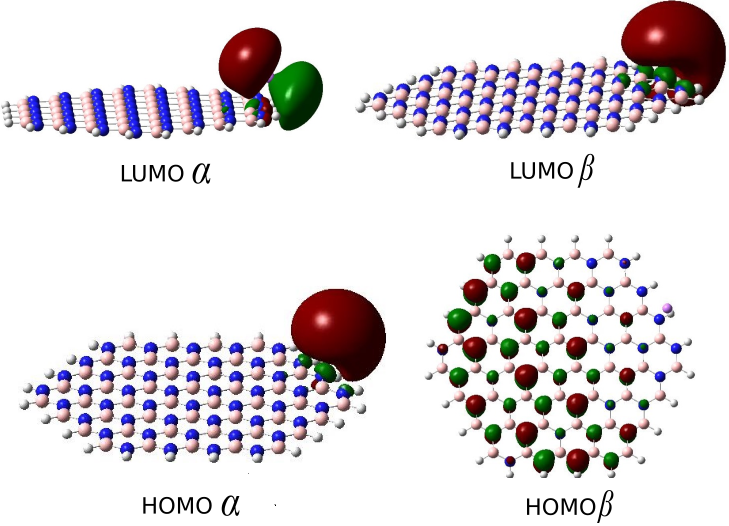}
   \caption [Models]{HOMO and LUMO molecular orbitals for spin up ($\alpha$) and spin down ($\beta$) electrons in a Li-doped BN flake (clusters No. 9-1, 9-2, 10 and 11-1). The red and green colors stand for the positive and negative signs of molecular orbital wave functions, respectively. The isovalue surface is 0.02 $e/(Bohr)^3$.} 
    \label{MO2}
    \end{figure}

About the frontier MOs, for pristine BN flakes, HOMO is not localized, and these MOs are concentrated at N atoms. The LUMO is concentrated on B atoms in the edge, and by the increase of the size, this concentration is more obvious (Fig. \ref{MO1}). Besides, by the increase of the size, the LUMO state has a blueshift, but the HOMO state has a redshift; consequently, the gap decreases as the size increases and the hardness of flakes are reduced \cite{li2017tuning}. Furthermore, HOMO states are located on N atoms and LUMO states for $B_{27} H_{18} N_{27}$ and $B_{36} H_{24} N_{36}$ are located on B atoms in the edges. So, by the increase of the size, the concentration of HOMO on N atoms and LUMO on the edge B atoms increase (Fig. \ref{MO1}). For doped BN flakes, HOMO states for spin-up electrons are localized $s$ orbitals that are concentrated on Li atoms, and minor molecular orbitals are scattered on neighbor atoms (Fig. \ref{MO2}, down left part). LUMO molecular orbitals for spin-up electrons are also localized on the edges. Theses orbitals for all clusters are localized $\pi$ orbitals concentrated on Li atoms and parallel to the BN surfaces (Fig. \ref{MO2}, top left part). The spin-down HOMO molecular orbitals are not localized; those are concentrated on some of N atoms away from the Li atom (Fig. \ref{MO2}, downright part). For spin up electrons, the main LUMO orbitals are also localized on the Li atom, and those are $s$ type molecular orbitals (Fig. \ref{MO2}, top right part).
On the other hand, we have mentioned that the adsorption of a Li atom makes the surface of BN to be insulator for spin-down and a semiconductor for spin-up electrons. Based on the above analysis, this point is understandable. For spin-up electrons, HOMO and LUMO orbitals have overlap. Hence, the $\alpha$ gap is low. Instead, for spin-down electrons, the overlap between a delocalized HOMO and a localized LUMO molecular orbital is less. So, for spin-down electrons, the gap is in the range of an insulator. Also, for spin-up electrons, these semiconductors have localized molecular orbitals; so, there is not any conduction as the result of this localized orbitals through the surface, and electrical isolation happens. This point is also notable for PL applications (we will discuss this point where we study applications of these flakes).

\subsection{Charge distributions and dipole moments}
In the case of $B_{12} H_{12} N_{12}$, the maximum negative charge belongs to some of the edge N atoms, and it is equal to -0.52 Mulliken. These atoms have neighbor H atoms with 0.24 Mulliken charges. The charge of the edge B atoms is 0.33 Mulliken, and the neighbor H atoms are negatively as much as -0.10 Mulliken charged. The maximum positively charged atoms are B atoms with 0.48 Mulliken. For $B_{27} H_{18} N_{27}$, the charge distribution is similar to the above structure. However, by the increase in size, the charge distribution interval has increased. The charge variation interval is from -0.57 Mulliken for some N atoms in the edges to 0.48 Mulliken for a B atom. For  $B_{36} H_{24} N_{36}$, this variation interval is from -0.54 for some edge N atoms to 0.49 Mulliken for B atoms.    

\begin{figure}[h]
   \centering   
   \includegraphics[scale=0.54]{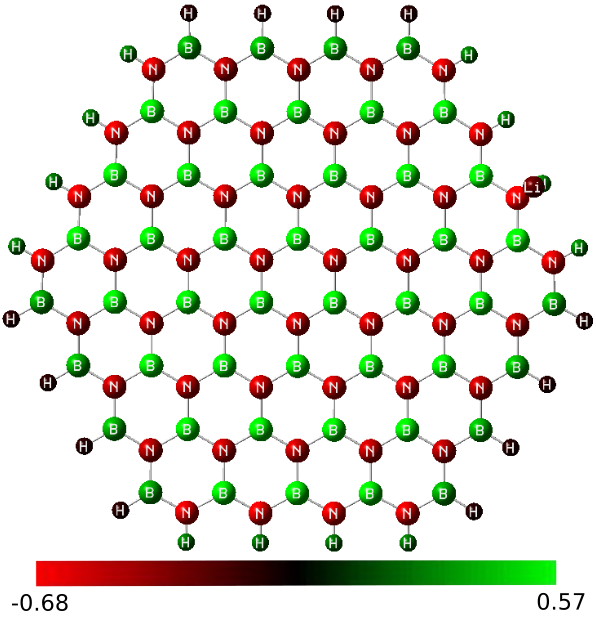}
   \caption [Models]{Charge distribution of flakes No. 9-1, 9-2, 10, and 11-1. The charge distribution interval changes from -0.68 to 0.57 Mulliken and the symbol of each atom is written on it.} 
    \label{4}
\end{figure}

According to Table. \ref{tab:table1}, in the case of doped BN flakes, Li atom is negatively charged in comparison to positively charged Li atoms in the case of graphene \cite{likhodam}. For clusters No. 1, 2, 4, 5, 6-1, 9-1, 9-2, 10, 11-1, and 12, surfaces have lost more than $20\%$ of their charge. These structures all have adsorbed a Li atom above a $N-H$ bond (Fig. \ref{4}). Clusters No. 6-2 and 11-2 that have adsorbed a Li atom near to the B decorated edge have lost respectively $11\%$ and $14\%$ of their charge. Other clusters, which means clusters No. 7, 8, and 3 that have a Li atom near to the center of the BN, have given less than $4\%$ of the electronic charge to the Li atom. The maximum negatively charged atom belongs to an N atom in the clusters No. 9-1, 9-2, 10, and 11-1. This negative charge amount is equal to -0.68 Mulliken charge. Clusters No. 6-1, 4, 5, 12, 1, and 2 have an N atom with a negative charge above -0.64 Mulliken. The adsorbed Li atom is near to the maximum negatively charged N atom in these structures. The most positive charge also belongs to a B atom, and that is 0.57 Mulliken. The highest positively charged B atom for other clusters is around 0.50 Mulliken.
Those H passivated B atoms have an average -0.10 Mulliken charge. However, H passivated N atoms are positively charged due to the different electronegativity. The electronegativity of an H atom is more than a B atom, and it is less than an N atom. The average charge of H atoms connected to the N atoms for clusters No. 1, 2, 4, 5, 6-1, 9-1, 9-2, 10, 11-1, and 12 is between 0.26 to 0.28 Mulliken. For other clusters, the H atom charge is 0.23 Mulliken. This amount is near to what we have in pristine BN flakes.   
    
About the dipole moment, the larger structures that have a Li atom adsorption near to a $N-H$ bond have greater polarity. According to Table. \ref{tab:table1}, the highest dipole moments belong to clusters No. 9-1, 9-2, 10, and 11-1. Then, clusters 12, 6-1, 4, 5, 1, and 2 have the highest dipole moment amounts, respectively. All of these clusters have adsorbed a Li atom above a $N-H$ bond. Clusters 11-2 and 6-2 that have a Li atom near to the B decorated edge have a dipole moment equal to 4.7 and 3.8 Debye, respectively. Other doped structures that have a Li atom near to the center (clusters No. 3, 7, and 8) have less than 1 Debye dipole moment.   

\subsection{Application suggestions}
The inorganic 2D layer of BN is attractive for clean energy applications. Additionally, the current Li-ion batteries should never be operated in a thermally insulated case or a high ambient temperature because of some problems such as fire, explosion, and thermal runaway that each one can lead to serious health and safety issues. Consequently, we suggest coating BN on current collectors which are aluminium (connected to the cathode) and copper (connected to the anode) flat sheets in a Li-ion battery \cite{yue20183d, ye2020ultralight} where the anode is the source of Li ions, the cathode is the sink for these ions, and the Li-ions flow through the electrolyte. This application guarantees the safety of the application of Li-ion batteries in different situations; for example, high temperature and pressure conditions because of the thermal management of rechargeable Li-ion batteries \cite{divakaran2020design}. BN nanosheets can sustain up to 850 $^{\circ}C$ \cite{o}; consequently, by the increase of the temperature, the structure is strong enough to tolerate the explosion of the battery. Its inertness beside the Li atoms and having localized molecular orbitals make the BN a good insulator, a protective coating, and a separator for Li atom environments. Hence, a layer of BN insulator increases Li-ion battery performance efficiency. According to the results of this article, the usage of this cover sheet does not affect reversible Li-ion battery performance, especially when BN is large enough to ignore the effects of the adsorption of a Li atom on the edge. We also predict that the increase of the temperature will cause that the central Li atoms go to the edge where the adsorption energy is maximum \cite{25likhodam}. So, a large area of the BN sheet is a good insulator for Li-ion applications. Consequently, we suggest the application of the BN sheet between the collector and a metal case. A metal case is quite useful to transfer the heat from the BN sheet to the environment. 

On the other hand, this wide gap material has interesting optical properties. According to what we have calculated, bandgap of pristine BN flakes, including $B_{12} H_{12} N_{12}$, $B_{27} H_{18} N_{27}$ and $B_{36} H_{24} N_{36}$, are $6.9$, $6.5$ and $6.4$ eV. These bandgap energies are in the range of deep ultraviolet luminescence ($h \nu \geq 5$ eV) \cite{prb}. These bandgap energies are also dependent on the size of the flakes. Moreover, by the adsorption of the Li atom, only the spin-down electrons can have such bandgap energy. Hence, spin-dependent photon emission is possible, what it has been done before by considering the external magnetic field \cite{exarhos2019magnetic}, and considering vacancies in BN sheet \cite{spin}, and such an effect in the case of vacant BN could be observed by magnetic resonance spectroscopy measurements \cite{spin}. For Li-doped BN, emissions change in the order of several Deci eV dependent on the place of the adsorption on the edge state because the HOMO-LUMO gap also changes in this range for different flakes. The size of the flakes also applies to the changes in the excitation energy. Hence, the BN flakes give such a possibility to study light-matter interaction at the nanoscale especially for nanophotonics. So, the adsorption makes it possible to manipulate and readout of spin defects in doped BN flakes. The Li-doped BN based PL is sensitive to the size and dopant place. This effect deduces to the multicolor fluorescent spin-based PL. We also predict that this PL effect could be possible to be observed in the laboratory by the time- and energy-resolved PL spectroscopy method \cite{prb}.  
\section{Conclusion}
We have studied the electronic properties of the pristine hydrogen-edged BN flakes and the interaction of the Li atom with these clusters. Accordingly, for pristine flakes, stability increases as the size increases. For doped flakes, the adsorption energy also increases as the size increases. And, for the $N-H$ bond adsorption, we have the maximum stability. The distance of the adsorbed Li atom from the BN surface decreases as the adsorption place becomes farther from the symmetry point of the BN flake. For pristine BN the bond lengths do not change as the size increases, and the changes of the bond lengths of doped BN are in order of 0.01 $\AA$. About the gap, for pristine BN, the gap increases as the size decreases. Furthermore, for doped structures, a little spin is transferred to the doped BN layer and some states are spin-polarized. A large value of the HOMO-LUMO gap for pristine and spin-down current leads to this point that we have a wide gap insulator for pristine BN and spin-down states. While, for spin-up electrons, the gap amounts are between 2 and 2.9 eV indicate that we have a semiconductor. Additionally, in doped cases, by the increase of the $\alpha$ gap, the $\beta$ gap of the spin-up states decreases. Frontier molecular orbitals have also been studied to discuss the ground and excited states. The trends seen in the DOS spectrum and charge distribution are also observed in the present work for pristine and Li-doped BN flakes. Moreover, the doped BN flakes possess non-zero dipole moments due to the asymmetry in the charge distribution. We have also suggested an application for the BN as a protecting layer for a Li-ion battery. Besides, its photoluminescence properties for optoelectronic applications have been discussed. This provides a good starting point for discussion and further research to study optical properties of Li-doped boron nitride and calculations of excited states of these flakes. Additionally, hydrogen-edged BN and related Li-doped flakes might prove an important area for future research in the field of luminescence properties for bioimaging. Moreover, further research is needed to confirm these novel findings for oxygen edged BN flakes that are common edge states in the exfoliation based produced BN flakes. 
\section{Acknowledgments}
This work was made possible by the facilities of Computational Nanotechnology Supercomputing Center, Institute for Research in Fundamental Science (IPM).
\bibliography{bib} 
\bibliographystyle{ieeetr}
\end{document}